\documentclass[twocolumn]{aastex62}

\usepackage{amsmath,amstext}
\usepackage[figure,figure*]{hypcap}

\received{receipt date}
\revised{revision date}
\accepted{acceptance date}
\published{published date}
\submitjournal{The Astrophysical Journal}

\shorttitle{Dissociation of the 1-Ethynylpyrene Cation}
\shortauthors{Rouill{\'e} et al.}

\begin{document}

\title{Threshold Dissociation of the 1-Ethynylpyrene Cation at Internal Energies Relevant to H~{\small I} Regions}

\author{Ga{\"e}l Rouill{\'e}}
% https://orcid.org/0000-0002-4016-1461
\affiliation{Laboratory Astrophysics Group of the Max Planck Institute for Astronomy at the Friedrich Schiller University Jena, Institute of Solid State Physics, Helmholtzweg 3, D-07743 Jena, Germany}

\author{Mathias Steglich}
% https://orcid.org/0000-0001-8206-6588
\affiliation{Laboratory for Synchrotron Radiation and Femtochemistry, Paul Scherrer Institute, CH-5232 Villigen, Switzerland}

\author{Patrick Hemberger}
% https://orcid.org/0000-0002-1251-4549
\affiliation{Laboratory for Synchrotron Radiation and Femtochemistry, Paul Scherrer Institute, CH-5232 Villigen, Switzerland}

\author{Cornelia J{\"a}ger}
\affiliation{Laboratory Astrophysics Group of the Max Planck Institute for Astronomy at the Friedrich Schiller University Jena, Institute of Solid State Physics, Helmholtzweg 3, D-07743 Jena, Germany}

\author{Thomas Henning}
% https://orcid.org/0000-0002-1493-300X
\affiliation{Max Planck Institute for Astronomy, K{\"o}nigstuhl 17, D-69117 Heidelberg, Germany}

\correspondingauthor{Cornelia J\"ager}
\email{cornelia.jaeger@uni-jena.de}

\begin{abstract}
Photoelectron photoion coincidence spectroscopy has been used to measure the threshold photoelectron spectrum of 1-ethynylpyrene and to obtain the breakdown graph describing the dissociation of the 1-ethynylpyrene cation. The threshold photoelectron measurement has allowed us to improve the determination of the ionization energy of 1-ethynylpyrene at 7.391 $\pm$ 0.005~eV. Concerning the main dissociation channels, the analysis of the breakdown graph has given 3.70 $\pm$ 0.60~eV as the activation energy for the loss of one H atom and 2.98 $\pm$ 1.80~eV for the loss of a second independent H atom. The corresponding entropies of activation are affected by large errors as observed in similar studies of other polycyclic aromatic hydrocarbon cations. Minor dissociation channels were also detected and identified as the loss of the C$_2$H group and the loss of a C$_2$H$_2$ unit and/or that of an H atom plus the C$_2$H group. The activation energies and the entropies of activation of these minor pathways could not be derived from the measurements. It is found that the cation of 1-ethynylpyrene behaves like the cation of pyrene and is consequently more photostable than the cation of 1-methylpyrene. We conclude that photodissociation is not the leading cause of the low abundance, if not the absence, of ethynyl-substituted polycyclic aromatic hydrocarbon species in the interstellar medium.
\end{abstract}


\section{INTRODUCTION}

The presence of neutral and ionized polycyclic aromatic hydrocarbon (PAH) molecules in the interstellar medium (ISM) is widely acknowledged \citep[for a review, see][]{Tielens08}. Indeed, their collective mid-IR emission spectrum explains best most of the features designated as the unidentified IR (UIR) emission bands; accordingly, the corresponding features are nowadays known as the aromatic IR bands (AIBs). Regular PAH molecules, however, cannot cause the emission bands observed around 3.4~$\mu$m wavelength. Similar to spectral features of aliphatic hydrocarbon species \citep[][and references therein]{Yang17}, the bands have been attributed to methylated and hydrogenated PAH molecules essentially \citep[][and references therein]{Maltseva18}.

Except for the proposed methylated and hydrogenated species, derivatives of PAH molecules do not appear to contribute to the UIRs. And yet the chemical network to which interstellar PAH molecules belong may include and link hydrocarbon substances of all sorts. For instance, a possible growth mechanism of PAH molecules in the ISM would give birth to ethynyl-substituted derivatives \citep{Mebel08}. Of course, the structural property that defines a category of molecules can enable its facile destruction, thus preventing the development of a population large enough to be detected.

The destruction of interstellar molecules proceeds mainly through photodissociation triggered by impinging VUV photons. For that reason, we have evaluated experimentally the photostability of two ethynyl-substituted PAH cations \citep{Rouille15}. We have concluded that ethynyl-substituted PAH species are as photostable as the corresponding regular species, whereas it had been previously observed that methylated and hydrogenated PAH molecules are less photostable \citep{Jochims99}. Double-imaging photoelectron photoion coincidence (i2PEPICO) spectroscopy was used in our previous study, albeit in a mode that did not allow us to analyze the breakdown graphs with the Rice-Ramsperger-Kassel-Marcus theory for unimolecular reactions \citep[RRKM theory;][]{Rice27,Kassel28,Marcus52}. As a consequence, the threshold activation energy ($E_0$) for the first dissociation of the cations and the corresponding entropy of activation ($\Delta^{\ddagger}S$) were not determined.

The recent report on the unimolecular dissociation of the 1-methylpyrene cation \citep{West18b} and the improved measurements on the pyrene cation \citep{West18a} motivate the present study. Although  they clearly differ in terms of photostability, they are both prototypes of AIB emitters. We have examined the dissociation of the 1-ethynylpyrene cation at threshold in order to compare it with the results reported for the pyrene and 1-methylpyrene cations. This comparison gives us information on the chemical network of interstellar PAH species.

\section{EXPERIMENTAL}

\begin{figure}
\epsscale{1.1}
\plotone{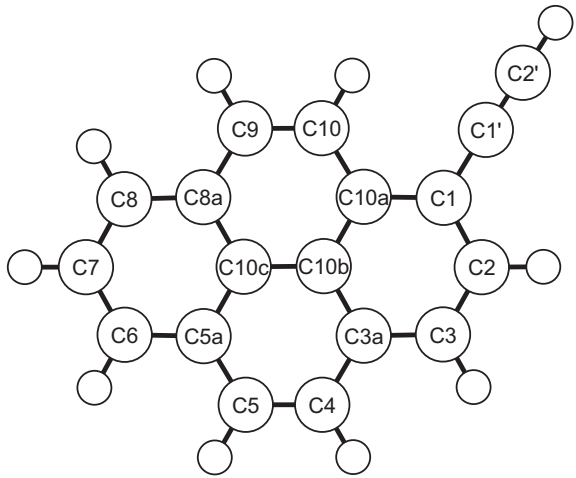}
\caption{Structure of 1-ethynylpyrene (C$_{18}$H$_{10}$) with numbering of the C atoms. \label{fig:fig1}}
\end{figure}

Figure~\ref{fig:fig1} shows the structure of 1-ethynylpyrene. The sample was purchased from abcr as a powder with a purity of 96{\%}. It was dispersed in a ball of glass wool placed in an oven equipped with an aperture 1~mm in diameter. This procedure facilitates the evaporation of the sample by creating a large surface area, which is critical when evaporation has to compete with heat-induced polymerization. Installed in a vacuum chamber coupled to that of the i2PEPICO spectrometer \citep{Bodi12,Sztaray17}, the oven was heated to 70~$^\circ$C during the experiments. An effusive beam of 1-ethynylpyrene molecules was thus produced, which was directed through a large aperture toward the ionization zone of the i2PEPICO spectrometer.

The threshold photoelectron (TPE) spectrum was derived from time-of-flight mass spectra measured for photon energies from 7.3 to 8.0~eV with a step interval of 0.005~eV. The resolution of the light source was 1500 and a magnesium fluoride (MgF$_2$) plate suppressed the harmonics of the synchrotron radiation. Energy calibration was carried out using Ar lines. The acquisition time for each mass spectrum was 30~s. The energy range was scanned twice and two TPE spectra were derived. They were added to each other and normalized against the energy-dependent photon flux that had been measured using a photodiode. An acceleration field of 160~V~cm$^{-1}$ was set to extract ions.

Breakdown graphs were obtained by measuring and analyzing time-of-flight spectra taken at photon energies from 14 to 21~eV with a step interval of 0.2~eV. The resolution of the synchrotron source was 4000 and harmonics of the synchrotron radiation were filtered out by using Ne gas. The ion signal was accumulated for 240~s at each energy position. Two scans were performed resulting in a total accumulation time of 480~s at each position. Measurements were carried out for two values of the acceleration field, i.e., with two different time scales, in order to establish whether this parameter affected measurably the detection of fragment ions.

\section{CALCULATIONS}

We used the Gaussian~09 and Gaussian~16 sets of programs \citep{Gaussian09,Gaussian16} to calculate the structures, rotational constants, harmonic vibrational frequencies, and electronic energies of the species relevant to the study. These data are necessary to perform RRKM modeling and to characterize dissociation pathways. Implementing density functional theory (DFT), the calculations were carried out using the B3LYP functional \citep{Becke88,Lee88,Becke93,Stephens94} together with the 6-311++G(d,p) basis set \citep{Frisch84}. The tight convergence criteria and ultrafine integration grid of the program were applied to optimize geometries. The ultrafine grid was also used to compute and verify reaction paths with intrinsic reaction coordinate calculations \citep{Fukui81}.

In order to derive vibrational frequencies, the theoretical harmonic vibrational frequencies are multiplied by a scaling factor. For PAH cations, the factor to be applied to harmonic frequencies computed at the B3LYP/6-311++G(d,p) level is 0.98 for all modes, with the exception of the CH stretching modes, for which it is 0.955 \citep{Vala09,Szczepanski11}. In this work, scaled vibrational frequencies are used in the RRKM analysis only. When exploring and describing theoretical dissociation pathways, the zero-point energies of the various ions and structures are computed by using their harmonic frequencies. The impact of this approximation is marginal considering the accuracy of the calculations.

To calculate the dissociation energies of the CH bonds, we took the energy of an H atom in its ground state as obtained with the DFT-B3LYP/6-311++G(d,p) model, i.e., 13.67~eV. At the same level of theory, we derived energies of 2084.87 and 2104.26~eV for the ground states of C$_2$H ($^2\Sigma^+$) and C$_2$H$_2$ ($^1\Sigma^+_g$), respectively, including the zero-point vibrational contributions. These values were used to determine the asymptotic dissociation energies for the loss of the corresponding units.

\section{RESULTS}

For convenience, the molecules pyrene, 1-methylpyrene, and 1-ethynylpyrene are denoted as Pyr, 1MePyr, and 1EtyPyr from now on.

\subsection{Ionization Energy}

\begin{figure}
\epsscale{1.1}
\plotone{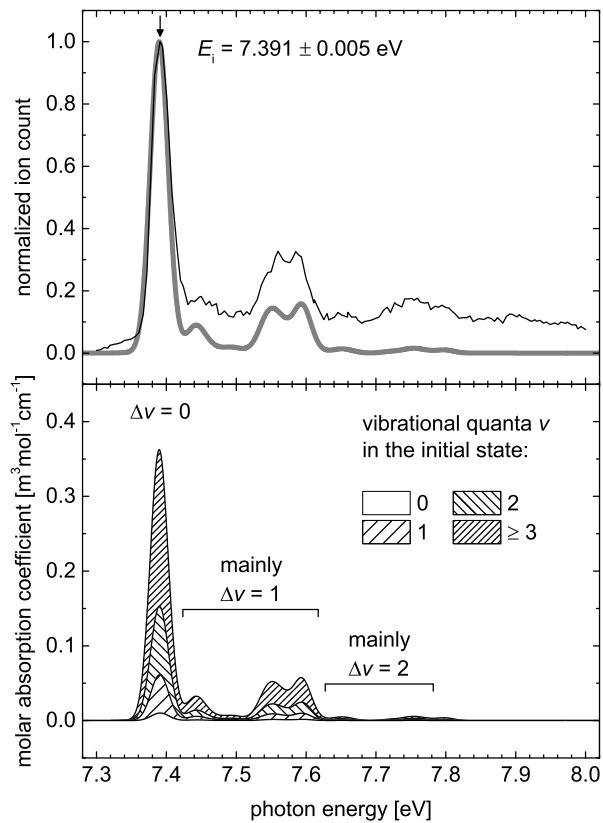}
\caption{Threshold photoelectron spectrum of 1EtyPyr. (Top:) The measured spectrum (solid black line) and a theoretical one (thick gray line). The arrow marks the measured ionization energy $E_\mathrm{i}$. (Bottom:) Analysis of the theoretical spectrum shown above in terms of quanta $v$ of vibrational excitation in the initial state and variation ${\Delta}v$ upon photon absorption. \label{fig:fig2}}
\end{figure}

Figure~\ref{fig:fig2} depicts the TPE spectrum of 1EtyPyr. The measured ionization energy, which corresponds to the position of the first ionization peak, is 7.391 $\pm$ 0.005~eV. It is in agreement with the earlier measurement at 7.41 $\pm$ 0.02~eV \citep{Rouille15}.

Figure~\ref{fig:fig2} compares the measured spectrum with a theoretical ionization spectrum. The origin of the latter, at 7.078~eV, has been shifted to the observed ionization energy value of 7.391~eV. The position of the bands in relation to the origin has not been scaled and a Gaussian function with a full width at half maximum of 240~cm$^{-1}$ is used as the common band profile. As the vibrational temperature of the molecules is set to 343~K, i.e., the temperature of the oven, the transitions originate mostly in vibrationally excited states. The bands arising on the high-energy side of the origin peak are caused by the superimposition of numerous components. The strongest bands, at 7.44~eV and in the 7.52--7.62~eV interval, correspond to the excitation of in-plane deformations of the carbon structure, both CC stretching and CCC bending modes. Hot bands expected on the low-energy side of the origin peak are too weak to be visible, in agreement with the measured spectrum. The resemblance between the measured and the simulated spectra supports the attribution of the signal at $m/z$ = 226 to 1EtyPyr$^+$.

\subsection{Fragment Ions}

\begin{figure}
\epsscale{1.1}
\plotone{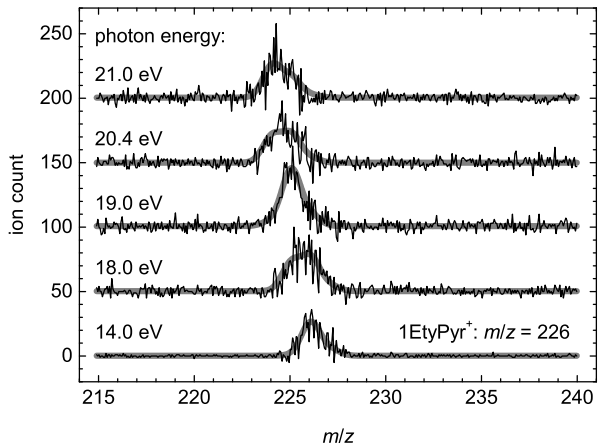}
\caption{Excerpts of mass spectra measured at key steps of the first photon-energy scan performed with an acceleration field of 80~V~cm$^{-1}$. The thick gray lines are fitted profiles described in Section~\ref{sec:breakdowngraph}. \label{fig:fig3}}
\end{figure}

The threshold time-of-flight mass spectra show three major peaks at $m/z$ values 224, 225, and 226. The peak at $m/z$ = 226 corresponds to the precursor ion 1EtyPyr$^+$ and the others are attributed to singly ($m/z$ = 225) and doubly ($m/z$ = 224) dehydrogenated fragment ions denoted [1EtyPyr $-$ H]$^+$ and [1EtyPyr $-$ 2H]$^+$, respectively. Arguments leading us to assign the double dehydrogenation to the separate emissions of H atoms rather than to the release of an H$_2$ molecule are given in Section~\ref{sec:paths}. Figure~\ref{fig:fig3} shows the peaks measured at specific photon energies, with 1EtyPyr$^+$, [1EtyPyr $-$ H]$^+$, and [1EtyPyr $-$ 2H]$^+$ appearing in turns as the main contributor to the mass spectrum. One notes that the resolution of the time-of-flight mass spectrometer is not sufficient to separate peaks differing by one unit in the $m/z$ scale.

\begin{figure}
\epsscale{1.1}
\plotone{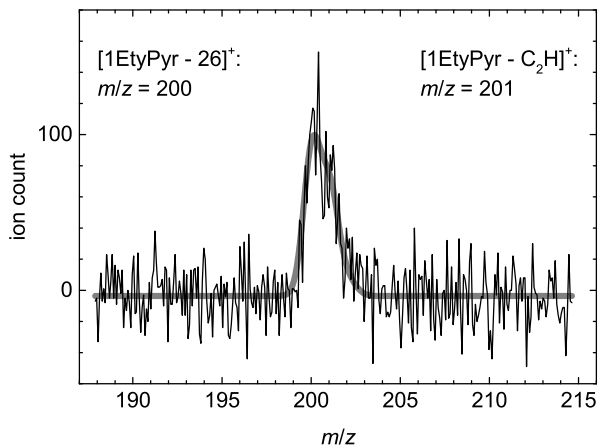}
\caption{Sum of the 16 mass spectra measured at photon energies of 19.6 to 21.0~eV during the two scans performed with an acceleration field of 160~V~cm$^{-1}$. The thick gray line is a fitted profile described in Section~\ref{sec:breakdowngraph}. \label{fig:fig4}}
\end{figure}

Two minor peaks were found at $m/z$ values 200 and 201 as demonstrated with Figure~\ref{fig:fig4}. The peak at $m/z$ = 201 is attributed to the fragment ion produced by the loss of the C$_2$H side group and denoted [1EtyPyr $-$ C$_2$H]$^+$. Ions with $m/z$ = 200 can be formed by the release of a C$_2$H$_2$ molecule or by the separate ejections of an H atom and the C$_2$H side group. The two channels are discussed in Section~\ref{sec:paths} and the fragment ions are denoted [1EtyPyr $-$ 26]$^+$ until their formation path is identified.

Interestingly, only [1EtyPyr $-$ H]$^+$ fragment ions were observed in our previous study \citep{Rouille15}. The present detection of supplementary fragments is attributed to a longer residence time of the excited ions in the acceleration zone, which is a consequence to using acceleration fields of 80 and 160~V~cm$^{-1}$ instead of 355~V~cm$^{-1}$.

\subsection{Breakdown Graph}\label{sec:breakdowngraph}

To build the breakdown graph, the relative fractions of precursor and fragment ions are derived for each photon energy value. It is done by comparing the areas of Gaussian functions fitted to the unresolved peaks arising in the mass spectra. The position of each Gaussian function is fixed during the fitting procedure and a common full width at half maximum, also fixed, is imposed to all profiles. Its value is 1.14 (resp. 1.04) for an acceleration field of 80~V~cm$^{-1}$ (resp. 160~V~cm$^{-1}$). Concerning each of the major ions 1EtyPyr$^+$, [1EtyPyr $-$ H]$^+$, and [1EtyPyr $-$ 2H]$^+$, a first Gaussian is used for the ions that contain only $^{12}$C atoms, and a second one for those comprising a $^{13}$C atom. The ratio of their areas is fixed to 0.1947, i.e., the relative abundance of $^{13}$C$^{12}$C$_{17}$H$_{10}$ compared to that of $^{12}$C$_{18}$H$_{10}$ when assuming the terrestrial natural isotopic abundances for C. The same procedure was applied to the minor fragment ions [1EtyPyr $-$ C$_2$H]$^+$ and [1EtyPyr $-$ 26]$^+$, with Gaussian profiles centered at the relevant $m/z$ values and given the full width at half maximum used for the major ions. The relative abundance of ions containing a $^{13}$C atom was changed to 0.1731.

\begin{figure}
\epsscale{1.1}
\plotone{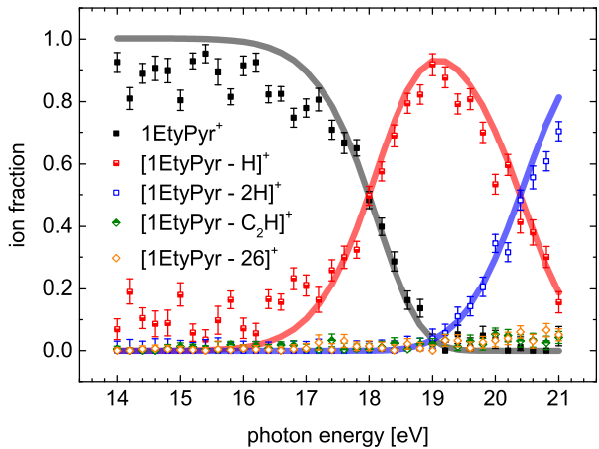}
\caption{Breakdown graph obtained for 1EtyPyr$^+$ with an acceleration field of 80~V~cm$^{-1}$. Each error bar represents the standard error resulting from the fit of the corresponding mass spectrum. The thick solid lines result from an RRKM analysis that neglects the minor dissociation channels. \label{fig:fig5}}
\end{figure}

The breakdown graph obtained from measurements with an acceleration field of 80~V~cm$^{-1}$ is presented in Figure~\ref{fig:fig5}. As expected from the peak heights in the mass spectra, the losses of one and two H atoms appear as the main dissociation channels of 1EtyPyr$^+$ in the energy range of the experiment. The dissociation mechanisms that generate the [1EtyPyr $-$ C$_2$H]$^+$ and [1EtyPyr $-$ 26]$^+$ ions are revealed as comparatively minor channels. A noisy background signal is observed between 14 and 17~eV at the mass of [1EtyPyr $-$ H]$^+$. This signal appears also in the measurements taken with an acceleration field of 160~V~cm$^{-1}$ (not shown), although it is then weaker and less noisy. Because the signal is constant until the rise indicating the formation of [1EtyPyr $-$ H]$^+$, and because we do not observe a corresponding creation of [1EtyPyr $-$ 2H]$^+$, we consider this signal to be an artifact.

\begin{figure}
\epsscale{1.1}
\plotone{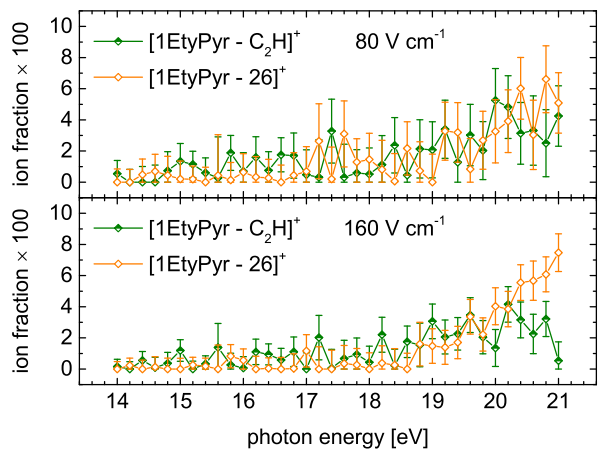}
\caption{The minor dissociation channels in the breakdown graph obtained for 1EtyPyr$^+$ with acceleration fields of 80~V~cm$^{-1}$ (top) and 160~V~cm$^{-1}$ (bottom). Each error bar represents the standard error resulting from the fit of the corresponding mass spectrum. \label{fig:fig6}}
\end{figure}

Figure~\ref{fig:fig6} shows the breakdown curves of the minor dissociation paths in detail, for the two acceleration fields. The curves suggest that the [1EtyPyr $-$ C$_2$H]$^+$ ions appear at an energy as low as 15~eV, while the rise of the [1EtyPyr $-$ 26]$^+$ ions starts at a higher energy, possibly around 17~eV. The number of ions transmitted to the detector increases with the acceleration field, hence the higher signal-to-noise ratio exhibited by the curves obtained with 160~V~cm$^{-1}$. They show that the fraction of [1EtyPyr $-$ C$_2$H]$^+$ ions reaches a maximum near 20~eV, whereas the fraction of [1EtyPyr $-$ 26]$^+$ ions keeps increasing with the photon energy.

\subsection{Theoretical Dissociation Paths}\label{sec:paths}

Prior to analyzing the breakdown graph, hypotheses as to the various dissociation paths it reveals have to be formulated, and information on the species and states that define these paths has to be obtained. In particular, given that the RRKM model makes use of a transition state to describe unimolecular reactions, the determination of the transition states that lead to dissociation may be critical. The rotational constants and vibrational frequencies of the species to be dissociated and those of the transition structures and states are determined by theoretical calculations.

The CH bond-dissociation energies in 1EtyPyr$^+$ have been computed and examined to determine which is involved in the loss of a single H atom. While there are only three different CH bonds in Pyr and Pyr$^+$ because of their high structural symmetry, all CH bonds are different in 1EtyPyr and its cation. For 1EtyPyr$^+$, the bond-dissociation energies derived from the calculations at the B3LYP/6-311++G(d,p) level of theory range from 5.17 to 5.42~eV for the CH bonds at aromatic rings; it is 5.46~eV for the CH bond of the ethynyl group. The CH bond of lowest dissociation energy (5.17~eV) is C7H, the middle bond of the only aromatic CH trio (Figure~\ref{fig:fig1}). Similarly, the middle bond in either of the two equivalent aromatic CH trios of Pyr$^+$ was reported as the one of lowest dissociation energy in that ion, with a theoretical value of 5.10~eV when computed with the B3LYP/6-311++G(d,p) model \citep{West14a}. Note that the ground state of 1EtyPyr$^+$ is a doublet state. To compute the CH bond-dissociation energies of this ion, we have taken the [1EtyPyr $-$ H]$^+$ fragments in their singlet state of lowest energy as we assume spin conservation throughout dissociation processes. Calculations show that the ground states of the singly dehydrogenated fragment ions are actually triplets, except when the missing H atom is the one in the ethynyl group.

The energy necessary for the dissociation of an aromatic CH bond in a PAH cation is higher than the barrier to the migration of an H atom and the formation of an aliphatic CH$_2$ group \citep[see examples in][]{Solano15,Trinquier17a,Castellanos18}. In 1EtyPyr$^+$, the energy for the dissociation of an aromatic CH bond is 5.17 to 5.42~eV (i.e., the bond-dissociation energy in the absence of a transition state, see the previous paragraph), while the barrier to H migration to the next C atom on the same ring is found to be 3.27 to 3.46~eV. As a consequence, the loss of an H atom can be preceded by the formation of a CH$_2$ group, from which the H atom is eventually released. The two dissociation routes link the same initial and final species and states, and in either case H loss occurs without the formation of a transition state. Thus, the routes do not differ in terms of dissociation energy, yet they are distinct as to the virtual transition state defined for the purpose of RRKM analysis. Since a number of vibrational modes of the transition state can be adjusted during this analysis, it is initiated by assuming that H loss happens through the dissociation of an aromatic CH bond.

The aromatic CH bonds in 1EtyPyr$^+$ do not differ greatly in energy according to the theoretical calculations, and the bond of lowest energy, C7H, has been selected to describe the loss of a single H atom. Because we have not found any transition state for H loss in 1EtyPyr$^+$, we have used the vibrational frequencies of the precursor ion as those of the transition state. A CH stretching mode is designated as a proxy for the one that would be associated with an imaginary vibrational frequency in the transition state \citep[see, e.g.,][]{Baer97,West18b}. The CH stretching mode that is missing after H loss at C7 is chosen for that role.

The loss of two H atoms has been treated as the successive losses of free atoms, i.e., the loss of an H atom by 1EtyPyr$^+$ followed by the loss of an H atom by [1EtyPyr $-$ H]$^+$. Taking into account the loss of free H atoms rather than the release of an H$_2$ molecule, direct or via the formation of a CH$_2$ group \citep{Solano15,Trinquier17a,Castellanos18}, is justified by the strong resemblance between the breakdown graphs obtained for 1EtyPyr$^+$ and Pyr$^+$. In the case of Pyr$^+$, the loss of two H atoms in the form of H$_2$ is a minor dissociation channel that may not be detected with iPEPICO \citep{West14a,West18a}, even though it was observed in a time-resolved photoionization mass spectrometry study \citep{Ling95a}. \citet{West14a} have actually demonstrated experimentally with the PIRENEA machine that the production of ions with the mass of [Pyr $-$ 2H]$^+$ proceeds through the formation of [Pyr $-$ H]$^+$ and results then from the loss of two independent H atoms. Additionally, \citet{Chen15} have estimated that H$_2$ release is not a significant dissociation channel for regular PAH molecules with internal energies not higher than 13.6~eV. According to that study, the dissociation of pyrene with formation of H$_2$ becomes a major mechanism when the internal energy reaches about 16.3~eV. Thus we infer that the release of an H$_2$ molecule is also a minor dissociation channel for 1EtyPyr$^+$.

Consequently bond-dissociation energies have been calculated for position isomers of [1EtyPyr $-$ H]$^+$. It is found that the CH bond of lowest dissociation energy is carried by the ring that misses an H atom, as in the case of dehydrogenated Pyr$^+$ \citep{West14a}. The bond-dissociation energy ranges then from 3.17 to 3.44~eV depending on the ring. As to the CH bond of the ethynyl group, it is more energetic, from 4.84 to 5.16~eV, depending on which other H atom is missing. Thus H atoms attached to the same carbon cycle are given a dominating role in the loss of two free H atoms. Because we have not found any transition state for the loss of an H atom in [1EtyPyr $-$ H]$^+$, the procedure described above for the loss of a single H atom in 1EtyPyr$^+$ has been repeated, with [1EtyPyr $-$ H]$^+$ in place of 1EtyPyr$^+$. The first and second dissociated bonds are C7H and C6H, respectively.

The loss of a C$_2$H unit is directly associated with the removal of the ethynyl group from 1EtyPyr$^+$. The theoretical bond-dissociation energy is 6.56~eV and the dissociation process by elongation of the bond between the C$_2$H group and the aromatic ring to which it is attached does not show any transition state. When comparing the zero-point vibrational energies of 1EtyPyr$^+$ with those of [Pyr $-$ C$_2$H]$^+$ and C$_2$H, a difference amounting to 2019~cm$^{-1}$ is found. Thus, the vibrational mode that features the stretching of the bond between the polycyclic structure and the side group appears, with a frequency of 2161~cm$^{-1}$, as a possible proxy for the imaginary vibrational frequency in the virtual transition state.

At least two dissociation paths can generate the [1EtyPyr $-$ 26]$^+$ ions. They are the loss of C$_2$H$_2$ by 1EtyPyr$^+$ and the loss of an H atom together with that of the C$_2$H group.

With regard to the loss of a C$_2$H$_2$ unit, pathways similar to those described by \citet{Trinquier17b} and \citet{West19} for Pyr$^+$ were verified. Like Pyr$^+$, 1EtyPyr$^+$ features direct routes for the loss of C$_2$H$_2$ from an inner ring or from an outer one \citep[respectively dissociation types 1 and 2 in][]{Trinquier17b}. It also features indirect routes via the formation of adjacent five- and seven-membered carbon rings \citep{West19}. Such isomerization was reported earlier in studies of the naphthalene cation \citep{Solano15,Bouwman16}. The routes that do not involve the cycle carrying the ethynyl group feature the same energy barriers as those computed for Pyr$^+$.

\begin{figure*}
\epsscale{1.1}
\plotone{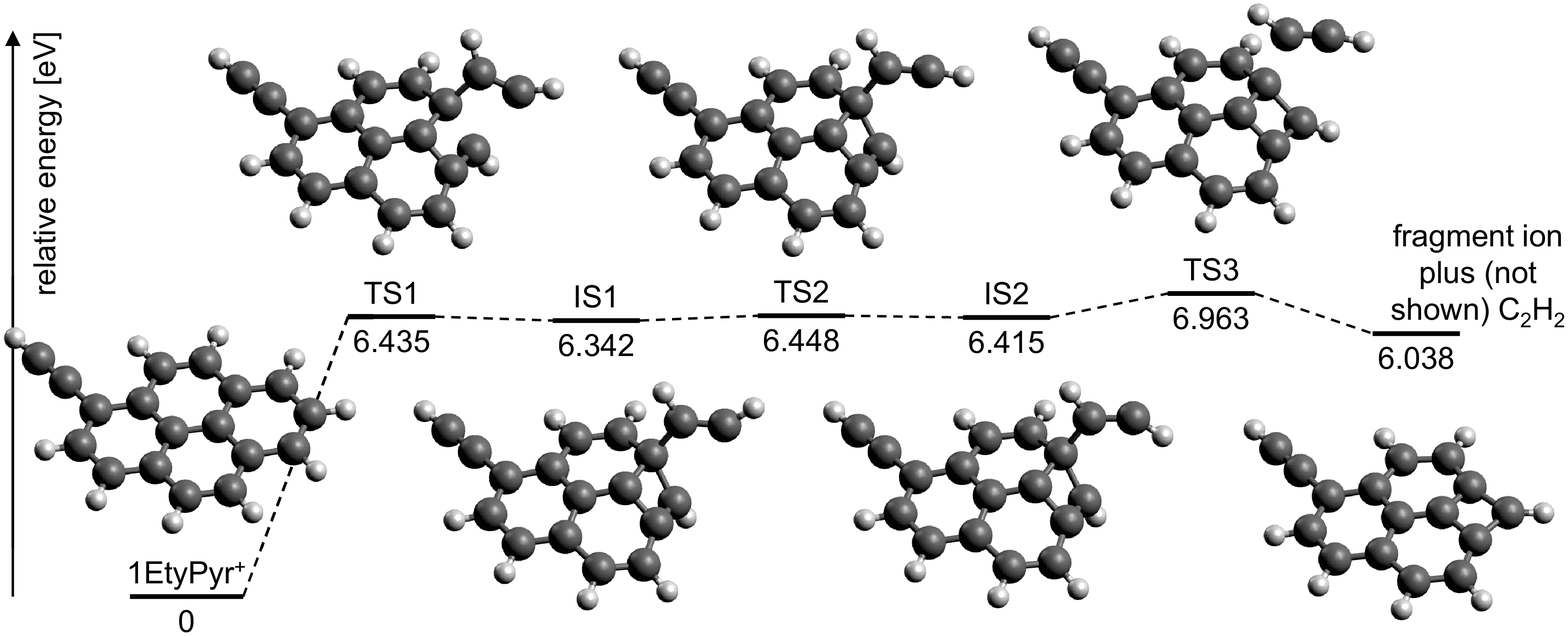}
\caption{Direct pathway of lowest energy barrier for the loss of C$_2$H$_2$ in 1EtyPyr$^+$. It is a direct path of type 2 and it features three transition states (TS1, TS2, and TS3) and two intermediate structures (IS1 and IS2), all chiral. The horizontal bars indicate the position of each structure on the vertical axis of relative energies. The energy values, printed under the bars, take into account the zero-point contributions. Bonds between atoms are only visual guides. \label{fig:fig7}}
\end{figure*}

As regards direct routes for C$_2$H$_2$ loss, four energetically distinct paths of type 1 were successfully computed and two of type 2. We have not found any direct pathway for the loss of C$_2$H$_2$ from the cycle carrying the ethynyl group. The loss of C$_2$H$_2$ proceeds through three transition states and two intermediate structures, whether the path is of type 1 or type 2. We have found that the energy barrier to the loss of C$_2$H$_2$ in 1EtyPyr$^+$ is lower via a path of type 2 (6.96~eV) than via a path of type 1 (7.40~eV). Figure~\ref{fig:fig7} depicts the path with the lowest energy barrier in which C7 and C8, and the H atoms they carry, are lost.

\begin{figure*}
\epsscale{1.1}
\plotone{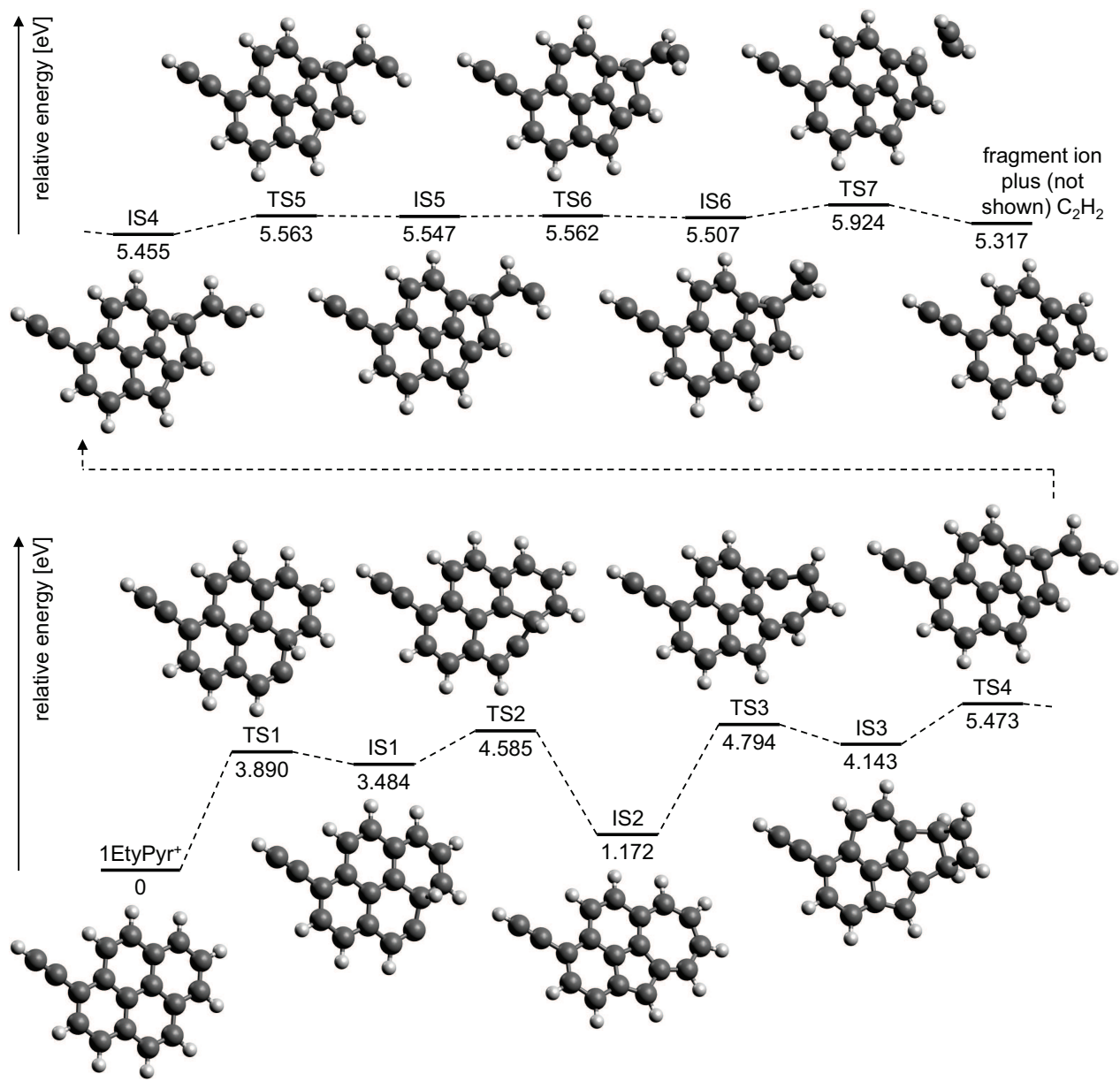}
\caption{Indirect pathway for the loss of C$_2$H$_2$ in 1EtyPyr$^+$. See caption of Figure~\ref{fig:fig7} for explanation. \label{fig:fig8}}
\end{figure*}

Two indirect paths leading to C$_2$H$_2$ loss were calculated following \citet{West19}. They feature energy barriers to dissociation of 5.92 and 5.95~eV, to be compared with 5.95~eV for Pyr$^+$ \citep[see Figure 1 in][]{West19}. The transition states were confirmed with IRC calculations, except in the case of TS6 for which this calculation did not lead to any solutions. Figure~\ref{fig:fig8} shows the indirect paths initiated by the migration of an H atom from C5 to C5a with the dissociation energy of 5.92~eV. The path initiated by H migration from C9 to C8a shows similar stages and energies, including a dissociation energy of 5.95~eV. A third indirect path starting with H migration from C4 to C3a is likely and would give a dissociation energy of 6.13~eV according to preliminary calculations.

Paths in which the ethynyl group is used to form the departing C$_2$H$_2$ unit have been looked for without success. The formation of C$_2$H$_2$ by direct association of the C$_2$H side group with the nearest H atom (at C2) could not be obtained. Additionally, in an attempt to form a C$_2$H$_2$ side chain indirectly, we found that the H atom at C2 can migrate to C1 and then to the side chain by getting over successive barriers that rise to 3.54 and 4.19~eV above the 1EtyPyr$^+$ ground energy, in this order. After passing the second barrier, however, the H atom does not stop at C1' and moves to the end of the side chain at C2', eventually forming a stable $\dbond$C$\dbond$CH$_2$ group.

As proposed above, the [1EtyPyr $-$ 26]$^+$ ions may be the products of the loss of an H atom and that of the C$_2$H group separately. This mechanism would cost a minimum of 9.84~eV, however, that is when the lost H atom is the one at C2, the closest to the C$_2$H group. Such energy is more than 2~eV higher than the energy barrier for the loss of a C$_2$H$_2$ unit through a direct or an indirect path, making the loss of a C$_2$H$_2$ unit the more likely source of [1EtyPyr $-$ 26]$^+$ ions in the lower energy range of the experiment. On the other hand, the competition between [1EtyPyr $-$ 26]$^+$ and [1EtyPyr $-$ 2H]$^+$ suggested by the experimental breakdown graphs would be consistent with the losses of an H atom and that of the C$_2$H group producing [1EtyPyr $-$ 26]$^+$ ions in the higher energy range of the experiment.

\subsection{RRKM Analysis}

Modeling breakdown curves often yields insight into the dissociative ionization mechanism \citep[e.g.,][]{Hemberger13} or can provide highly accurate thermochemical values such as bond dissociation energies and heats of formation \citep[e.g.,][]{Bodi14,Rowland15,Voronova18}. An analysis of the breakdown graph according to the RRKM model has been performed with \citet{Sztaray10}'s program minipepico (version 194i). Only the channels involving H loss have been taken into account because the inclusion of two minor channels, notwithstanding the ambiguity concerning the formation of [1EtyPyr $-$ 26]$^+$, prevented convergence toward a complete solution.

The theoretical breakdown curves were fitted by taking into account the dissociation path of lowest energy for each type of loss (Section~\ref{sec:paths}), and an adequate degeneracy value to allow for the number of energetically equivalent paths. Thus the loss of a single H atom and that of two free H atoms are represented respectively by the loss of the H atom on C7 (Figure~\ref{fig:fig1}), and the loss of the H atoms on C6 and C7. Regarding the degeneracy value attributed to the two types of dissociation, we have chosen 10 for the loss of a single H atom and 1 for the loss of a second H atom.

As to the degeneracy of the single-H-loss channel, a value of 10 is chosen because the bond-dissociation energies of the ten CH bonds are similar as reported in Section~\ref{sec:paths}. Nine of the computed [1EtyPyr $-$ H]$^+$ structures are nonplanar in their lowest singlet state as the removal of a single H atom from an aromatic ring in 1EtyPyr$^+$ causes an out-of-plane deformation of this ring. Although the consequent chirality could set the degeneracy of the single-H-loss channel to 19, we ignore the deformation because we have not demonstrated that it is related to the structure of a transition state in the dissociation path.

Similarly, several of the doubly dehydrogenated ions are found to be nonplanar. We do not take the phenomenon into account and set the degeneracy for the loss of a second H atom to 1. Indeed the identity of the first lost H atom determines that of the second one as the bond-dissociation energy is the lowest for an H atom at an aromatic ring that already misses one. We neglect the fact that there are actually two options after the loss of the H atom at C7 since it has two neighbors as the middle member of a trio. We also disregard the higher energy necessary for losing a second H atom when the first one is the acetylenic H atom (H at C2').

The results of the analysis are given in Table~\ref{tbl:tbl1}, together with those obtained by other authors for Pyr$^+$ and 1MePyr$^+$. The header 2H loss refers to the loss of the second of two free H atoms. A temperature of 343~K, that of the source, has been assumed for the 1EtyPyr vapor.

\begin{deluxetable*}{llllll}
\tablecaption{RRKM Analysis\tablenotemark{a}\label{tbl:tbl1}}
\tablehead{ & \multicolumn{2}{l}{H loss} & \multicolumn{2}{l}{2H loss} & \\
\colhead{Species} & \colhead{$E_0$} & \colhead{$\Delta^{\ddagger}S_\mathrm{1000~K}$} & \colhead{$E_0$} & \colhead{$\Delta^{\ddagger}S_\mathrm{1000~K}$} & \colhead{Reference} }
\startdata
  1EtyPyr$^+$ & 3.70 $\pm$ 0.60 & -1 $\pm$ 34       & 2.98 $\pm$ 1.80 & -1 $\pm$ 110      & this work\tablenotemark{b} \\   % 80 fixed
              & 3.75 $\pm$ 0.52 & -1 $\pm$ 29       & 3.02 $\pm$ 1.50 & -1 $\pm$ 91       & this work\tablenotemark{c} \\   % 160 fixed
              & 3.74 $\pm$ 0.49 & -1 $\pm$ 28       & 3.02 $\pm$ 1.38 & -1 $\pm$ 82       & this work\tablenotemark{d} \\   % 160 free
  1MePyr$^+$  & 2.78 $\pm$ 0.25 & 6 $\pm$ 19        &                 &                   & 1 \\
  Pyr$^+$     & 4.16 $\pm$ 0.69 & -5 $\pm$ 37       & 3.09 $\pm$ 1.22 & -14 $\pm$ 84      & 2 \\
              & 5.4 $\pm$ 1.2   & 57 $\pm$ 89       & 3.3 $\pm$ 1.1   & 16 $\pm$ 84       & 3 \\
              & 4.60            & 10.7              & 4.10            & 13.3              & 4 \\
\enddata
\tablenotetext{a}{The values of $E_0$ are expressed in eV and those of $\Delta^{\ddagger}S_\mathrm{1000~K}$ are given in J~K$^{-1}$~mol$^{-1}$.}
\tablenotetext{b}{Field of 80~V~cm$^{-1}$, fixed frequencies.}
\tablenotetext{c}{Field of 160~V~cm$^{-1}$, fixed frequencies.}
\tablenotetext{d}{Field of 160~V~cm$^{-1}$, free frequencies.}
\tablerefs{(1) \citet{West18b}; (2) \citet{West18a}; (3) \citet{West14a}; (4) \citet{Ling95a}.}
\end{deluxetable*}

Table~\ref{tbl:tbl1} presents values of $E_0$ and $\Delta^{\ddagger}S_\mathrm{1000~K}$ obtained with acceleration fields of 80 and 160~V~cm$^{-1}$. The activation energies were determined by using the fitting procedure implemented in minipepico. In the experiment with a field of 80~V~cm$^{-1}$, the activation entropy values were not fitted, they were computed while keeping the vibrational frequencies of the transition structures or proxies fixed to the scaled theoretical values. We actually attempted to fit the $\Delta^{\ddagger}S_\mathrm{1000~K}$ values by using the five lowest vibrational frequencies of the transition structures as free parameters \citep{Baer97}. The attempts, however, lead to too high frequencies for the transition structure of the H-loss channel because of the background signal observed between 14 and 17~eV at the mass of [1EtyPyr $-$ H]$^+$. In contrast, it was possible to fit the activation entropy values from the measurements taken with an acceleration field of 160~V~cm$^{-1}$. The agreement between the results obtained with fixed and with free frequencies when treating the data for 160~V~cm$^{-1}$ (Table~\ref{tbl:tbl1}) validates the analysis of the measurements at 80~V~cm$^{-1}$ with fixed frequencies. Standard deviations were derived for the two sets of activation energy and entropy values with the procedure described in Section~\ref{sec:errors}. Note that the quality of the activation entropy values is evaluated whether the vibrational frequencies are free or fixed during the fit. In Table~\ref{tbl:tbl1}, the errors associated with the $E_0$ and $\Delta^{\ddagger}S_\mathrm{1000~K}$ values given for 1EtyPyr$^+$ represent each three times the standard deviation. Their values are similar in magnitude to those obtained by \citet{West18a} for Pyr$^+$ and for other PAH cations.

\section{DISCUSSION}

Despite a shift of $\sim$0.5~eV toward lower energies, the breakdown curves obtained for 1EtyPyr$^+$ closely resemble those measured for Pyr$^+$ \citep{West14a,West18a}. This means that the existence of the ethynyl group in 1EtyPyr$^+$ does not facilitate the loss of a C-containing unit. Not to be forgotten, the ionization energy of 1EtyPyr (7.391 $\pm$ 0.005~eV) is similar to that of Pyr \citep[7.415 $\pm$ 0.010~eV,][]{Mayer11}. In total, these facts are consistent with 1EtyPyr$^+$ behaving like Pyr$^+$ in terms of dissociation. This is valid for the low internal energies produced in the present experiment, which are relevant to the H~{\small I} regions of the ISM. Indeed internal energies ranging from 6.6 to 13.6~eV are deposited in the precursor ions, as obtained after subtracting the ionization energy of 1EtyPyr from the photon energy.

The theoretical dissociation energies of the aromatic CH bonds are almost identical for 1EtyPyr$^+$ and Pyr$^+$. In 1EtyPyr$^+$, the dissociation energy of the acetylenic CH bond is slightly higher than that of any aromatic CH bond, and the difference strongly increases in the singly dehydrogenated cation. If the prevailing H-loss channel involved the H atom of the side chain, the contribution of [1EtyPyr $-$ 2H]$^+$ to the breakdown curves would rise at a higher energy. Accordingly, it is inferred that 1EtyPyr$^+$ loses dominantly H atoms attached to aromatic carbon rings rather than the H atom of the ethynyl group. Thus the quantum chemistry calculations also support the resemblance between 1EtyPyr$^+$ and Pyr$^+$. This is in stark contrast to the behavior of 1MePyr$^+$, for which the first dissociation is the loss of an H atom belonging to the methyl group rather than an H atom attached to an aromatic carbon ring \citep{West18b}. The difference is reflected in the breakdown curves of this cation, which are strongly shifted ($\sim$4~eV) toward lower energies with respect to those of Pyr$^+$. It shows also in its ionization energy \citep[7.27 $\pm$ 0.01~eV,][]{West18b}, which is clearly lower than those of 1EtyPyr and Pyr.

Since 1EtyPyr$^+$ and Pyr$^+$ behave similarly with respect to ionization and dissociation, we conclude that 1EtyPyr is as photostable as Pyr. Comparing the behaviors of 1MePyr$^+$ and Pyr$^+$, we find that 1MePyr is less photostable than Pyr, and consequently as 1EtyPyr. In the absence of obvious fundamental obstacles, we extend this result to PAH molecules of similar or larger sizes. Accordingly, PAH molecules carrying an ethynyl side chain are in general more photostable than those carrying a methyl group.

\begin{figure}
\epsscale{1.1}
\plotone{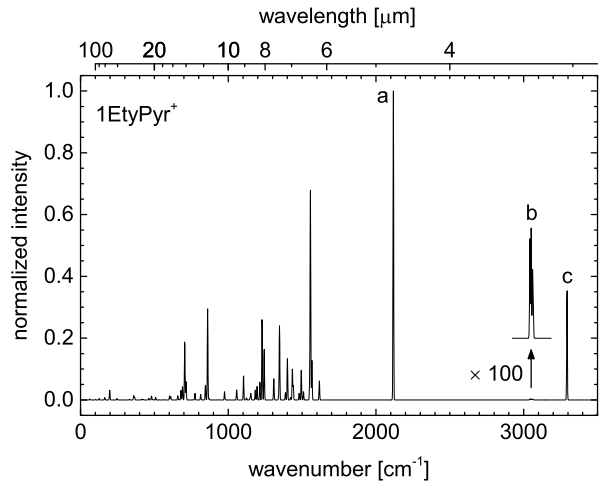}
\caption{Theoretical IR spectrum of 1EtyPyr$^+$. The vibrational frequencies are scaled as described in the main text and each peak is given a Gaussian profile with a full width at half maximum of 5~cm$^{-1}$. Latin characters indicate the bands corresponding to the CC stretching mode of the side chain (a), the aromatic CH stretching modes (b), and the acetylenic CH stretching mode (c). \label{fig:fig9}}
\end{figure}

It is assumed that interstellar methylated PAH molecules contribute to the infrared emission feature observed at 3.4~$\mu$m. In contrast, ethynyl-substituted PAH molecules have not been unequivocally linked to infrared emission bands. Actually, a feature at 15.8~$\mu$m was found to coincide with the CCH bending of a $\sbond$C$\tbond$CH structure \citep{Sloan14}. It must not be confused with a band at the same wavelength assigned to large neutral PAH molecules \citep{Peeters17}. The type of species it belonged to, however, whether PAH derivatives or aliphatic chains, was not determined. Nevertheless, compounded absorption spectra of neutral ethynyl- and butadiynyl-substituted PAH molecules do exhibit a band at 15.9~$\mu$m, attributed to the overlapping contributions by the in- and out-of-plane C$\tbond$CH bending vibrations \citep{Rouille12}. These compounded spectra also show a strong band at $\sim$3.03~$\mu$m that indicates nearly coincident acetylenic CH stretching frequencies. Limiting our examination to ethynyl derivatives, the C$\tbond$C stretching vibration gives rise to a very weak absorption at 4.77~$\mu$m. In the spectrum of cations, however, the very same vibration is responsible for a strong absorption. It corresponds to the strongest band in the theoretical IR spectrum of 1EtyPyr$^+$ displayed in Figure~\ref{fig:fig9}, at a theoretical position of 4.72~$\mu$m. Indeed, the neutral and positively-charged derivatives differ as to the charge distribution in the ethynyl chain.

To our knowledge, the UIR emission bands do not comprise members at $\sim$3.03 and/or $\sim$4.77~$\mu$m. Consequently the abundance of ethynyl-substituted PAH molecules and cations in the regions that show AIBs is below the current detection limit. The mechanism that populates the ISM with carriers of the AIBs and 3.4~$\mu$m feature is not well known and it is likely plural \citep[see discussion in][]{Galliano18}. Carriers may emanate from stellar sources, yet they may also be produced in molecular clouds. In both stellar and interstellar environments, simple chemical reactions can contribute to the formation of PAH molecules, and ethynyl-substituted derivatives are intermediates in either case. They appear in the hydrogen-abstraction-C$_2$H$_2$-addition mechanism thought to proceed in stellar environments \citep{Frenklach85,Frenklach89,Wang94}, and also in the ethynyl-addition mechanism that could take place in the ISM \citep{Mebel08}. Further details on the relevance of ethynyl-substituted PAH molecules to astrophysics have been presented earlier \citep{Rouille12}. Thus ethynyl-substituted derivatives could exist wherever PAH molecules, i.e., the carriers of the AIBs and 3.4~$\mu$m feature, are found. As they are not detected, an efficient destruction mechanism of these derivatives is at work, which is not photodissociation. Chemical reactions that involve the unsaturated ethynyl group can be proposed as an alternative.

The fact that ethynyl-substituted PAH molecules are intermediates in gas-phase mechanisms that form PAH molecules means that they are destroyed in these processes. Additional reactions could be looked for, especially where the cationic form of the ethynyl-substituted derivatives is involved. Finally, reactions in the solid phase are critical if the largest fraction of cosmic PAH substances goes through a solid-state stage. Indeed the carriers of the AIBs and 3.4~$\mu$m feature may be born from the evaporation of very small grains \citep{Rapacioli05,Pilleri15}. We know that ethynyl-substituted PAH derivatives can survive in soot grains synthesized in the laboratory \citep{Rouille12}. In cosmic grains, however, the derivatives may be destroyed by irradiation-induced reactions that are facilitated by the close proximity of reactants.

\section{CONCLUSIONS}

This study confirms that the cation of 1-ethynylpyrene (1EtyPyr$^+$) behaves like that of pyrene (Pyr$^+$) with regard to dissociation when taking into account internal energies up to 13.6~eV. Consequently 1EtyPyr$^+$ is more stable than the cation of 1-methylpyrene (1MePyr$^+$). Inferring that ethynyl-substituted PAH molecules and ions are in general more stable than methyl-substituted ones, it is interesting that only the latter category is observed in the ISM alongside the regular species. This suggests that the physical-chemical network to which interstellar PAH species belong does not favor the existence of ethynyl-substituted derivatives. Assuming that the interstellar PAH molecules to which the AIBs and 3.4~$\mu$m feature are attributed are mainly born from the evaporation of very small grains, we conclude that this mechanism does not produce ethynyl-substituted PAH substances in detectable amounts.

\acknowledgments
This work was carried out within the framework of a cooperation between the Max-Planck-Institut f{\"u}r Astronomie, Heidelberg, and the Friedrich-Schiller-Universit{\"a}t Jena. The support of the Deutsche Forschungsgemeinschaft (DFG) through project No. 281937660 is gratefully acknowledged by C.J and G.R. The PEPICO measurements were carried out under project No. 20171598 at the VUV beamline of the Swiss Light Source, located at Paul Scherrer Institute, Villigen, Switzerland. M.S and P.H gratefully acknowledge funding by the Swiss Federal Office of Energy (SI/501269-01) and technical support by Patrick Ascher. Finally, the authors express their thanks to the anonymous reviewers for their valuable comments.

\bibliographystyle{aasjournal}
\bibliography{roui1903rev1}

\appendix

\section{EVALUATION OF ERRORS IN RRKM ANALYSIS}\label{sec:errors}

The method described by \citet{Shuman08} to evaluate errors proved impractical for the same reason frequencies could not be used as free parameters, and also because it cannot be applied to a signal that is a minor contributor to the overall error as the one at the mass of [1EtyPyr $-$ C$_2$H$_2$]$^+$ is. Consequently, taking the solution obtained with minipepico, we computed standard deviations by performing what amounts to a single iteration of a least-squares fitting procedure for nonlinear functions. It consists of solving the system of equations
\begin{equation}
f_{i,j}^{\mathrm{o}}-f_{i,j}^{\mathrm{c}}(p_k)=\sum_k{\frac{f_{i,j}^{\mathrm{c}}(p_k+\Delta p_k)-f_{i,j}^{\mathrm{c}}(p_k)}{\Delta p_k}x_k} ,
\end{equation}
where $f_{i,j}^{\mathrm{o}}$ and $f_{i,j}^{\mathrm{c}}$ are respectively the observed and calculated ion fractions at energy position number $i$ for ion number $j$; $p_k$ is the value of parameter number $k$, i.e., one of the $E_0$ and $\Delta^{\ddagger}S_\mathrm{1000~K}$ values in Table~\ref{tbl:tbl1}, and $\Delta p_k$ is a variation applied to $p_k$. The standard deviation determined for unknown number $k$, i.e., $x_k$, is the standard deviation attributed to $p_k$.

Variation $\Delta p_k$ must be large enough to produce nonzero values of the approximate numerical derivative in a meaningful number of equations. Varying the activation entropy is accomplished by modifying the lowest vibrational frequency of the relevant transition structure. The variations we have used consist of an increase of 1{\%} for all parameters. Each $f_{i,j}^{\mathrm{c}}(p_k+\Delta p_k)$ is computed by using minipepico with all parameters fixed while replacing $p_k$ with $p_k+\Delta p_k$.

\end{document}